\documentclass[prd,aps,twocolumn,floatfix,superscriptaddress]{revtex4}

\usepackage{amssymb,graphicx}
\usepackage{epsfig}
\usepackage[usenames]{color}

\def\jnl@style{\it}
\def\aaref@jnl#1{{\jnl@style#1}}

\def\aaref@jnl#1{{\jnl@style#1}}

\def\aj{\aaref@jnl{AJ}}                   
\def\araa{\aaref@jnl{ARA\&A}}             
\def\apj{\aaref@jnl{ApJ}}                 
\def\apjl{\aaref@jnl{ApJ}}                
\def\apjs{\aaref@jnl{ApJS}}               
\def\ao{\aaref@jnl{Appl.~Opt.}}           
\def\apss{\aaref@jnl{Ap\&SS}}             
\def\aap{\aaref@jnl{A\&A}}                
\def\aapr{\aaref@jnl{A\&A~Rev.}}          
\def\aaps{\aaref@jnl{A\&AS}}              
\def\azh{\aaref@jnl{AZh}}                 
\def\baas{\aaref@jnl{BAAS}}               
\def\jrasc{\aaref@jnl{JRASC}}             
\def\memras{\aaref@jnl{MmRAS}}            
\def\mnras{\aaref@jnl{MNRAS}}             
\def\pra{\aaref@jnl{Phys.~Rev.~A}}        
\def\prb{\aaref@jnl{Phys.~Rev.~B}}        
\def\prc{\aaref@jnl{Phys.~Rev.~C}}        
\def\prd{\aaref@jnl{Phys.~Rev.~D}}        
\def\pre{\aaref@jnl{Phys.~Rev.~E}}        
\def\prl{\aaref@jnl{Phys.~Rev.~Lett.}}    
\def\pasp{\aaref@jnl{PASP}}               
\def\pasj{\aaref@jnl{PASJ}}               
\def\qjras{\aaref@jnl{QJRAS}}             
\def\skytel{\aaref@jnl{S\&T}}             
\def\solphys{\aaref@jnl{Sol.~Phys.}}      
\def\sovast{\aaref@jnl{Soviet~Ast.}}      
\def\ssr{\aaref@jnl{Space~Sci.~Rev.}}     
\def\zap{\aaref@jnl{ZAp}}                 
\def\nat{\aaref@jnl{Nature}}              
\def\iaucirc{\aaref@jnl{IAU~Circ.}}       
\def\aplett{\aaref@jnl{Astrophys.~Lett.}} 
\def\apspr{\aaref@jnl{Astrophys.~Space~Phys.~Res.}}
\def\bain{\aaref@jnl{Bull.~Astron.~Inst.~Netherlands}} 
\def\fcp{\aaref@jnl{Fund.~Cosmic~Phys.}}  
\def\gca{\aaref@jnl{Geochim.~Cosmochim.~Acta}}   
\def\grl{\aaref@jnl{Geophys.~Res.~Lett.}} 
\def\jcp{\aaref@jnl{J.~Chem.~Phys.}}      
\def\jgr{\aaref@jnl{J.~Geophys.~Res.}}    
\def\jqsrt{\aaref@jnl{J.~Quant.~Spec.~Radiat.~Transf.}}
\def\memsai{\aaref@jnl{Mem.~Soc.~Astron.~Italiana}}
\def\nphysa{\aaref@jnl{Nucl.~Phys.~A}}   
\def\physrep{\aaref@jnl{Phys.~Rep.}}   
\def\physscr{\aaref@jnl{Phys.~Scr}}   
\def\planss{\aaref@jnl{Planet.~Space~Sci.}}   
\def\procspie{\aaref@jnl{Proc.~SPIE}}   

\def\magstar{{\tt Magstar}}
\def\lorene{{\sc Lorene}}
\def\gigatesla{{\sc GT}}
\def\rmd{{\rm d}}

\def\be {\begin{equation}}
\def\ee {\end{equation}  }
\def\beq{\begin{eqnarray}}
\def\eeq{\end{eqnarray}  }

\font\bfgreek=cmmib10

\newcommand{\had}{{\sc had}}

\def\bbf{{\hbox{\bfgreek\char'146}}}

\def\bbq{{\hbox{\bfgreek\char'161}}}

\def\bbs{{\hbox{\bfgreek\char'163}}}

\def\bbw{{\hbox{\bfgreek\char'167}}}

\begin{document}

\title{Evolutions of  Magnetized and Rotating Neutron Stars}

\author{Steven L. Liebling}
\affiliation{Department of Physics, Long Island University -- C.W. Post Campus,
Brookville, NY 11548}
\author{Luis Lehner}
\affiliation{
 Perimeter Institute for Theoretical Physics,
 Waterloo, Ontario, Canada
}
\affiliation{
 Department of Physics,
 University of Guelph,
 Guelph, Ontario, Canada
}
\affiliation{
Canadian Institute for Advanced Research, Cosmology \& Gravity Program
}
\author{David Neilsen}
\affiliation{Department of Physics and Astronomy, Brigham Young
University, Provo, UT 84602}
\author{Carlos Palenzuela} 
\affiliation{
Canadian Institute for Theoretical Astrophysics (CITA),
 Toronto, Ontario, Canada
}
\affiliation{Max-Planck-Institut f\" ur Gravitationsphysik,
Albert-Einstein-Institut, Golm, Germany
}

\date{\today}

%
%
\begin{abstract}
We study the evolution of magnetized and rigidly rotating neutron stars
within a fully general relativistic implementation of ideal magnetohydrodynamics
with no assumed symmetries in three spatial dimensions.
The stars are modeled as rotating, magnetized polytropic stars and we examine diverse scenarios
to study their dynamics and stability properties. In particular we concentrate on the
stability of the stars and possible critical behavior.
In addition to
their intrinsic physical significance, we use these evolutions
as further tests of our implementation which incorporates new developments to handle magnetized
systems.
\end{abstract}

\maketitle

%
%
\section{Introduction}

Neutron stars play a key role in some of the most interesting  astrophysical
events observed, from supernova remnants and pulsars to a less certain role
in long gamma ray bursts (GRBs). As such they have attracted significant
research into their dynamics (for a recent review see~\cite{Stergioulas:2003yp}).
In this paper, we revisit the dynamics of rotating, and possibly magnetized,
neutron stars modeled as polytropic stars within a fully nonlinear, general
relativistic model of ideal magnetohydrodynamics (GR+MHD). We study the stability properties
of these stars and highlight possible critical behavior exhibited by the system.
Furthermore, our studies serve to demonstrate the effectiveness of  our code and certain
new developments discussed here. This code has been applied to
other astrophysical
problems~\cite{Palenzuela:2006wp,Palenzuela:2007dm,Anderson:2007kz,Anderson:2008zp,Palenzuela:2009yr,Megevand:2009yx}. 

In recent years, a number of fully relativistic evolutions (as opposed to those using Newtonian
gravity or other approximations to general relativity) of rotating
polytropes have appeared studying
gravitational wave production and black hole formation in 
astrophysically relevant systems. To this end,
studies beyond spherical symmetry are required which are computationally more
demanding. For instance, rotating stars require moving beyond
spherical symmetry and many interesting axisymmetric scenarios can be addressed using 2D implementations
allowing for excellent resolution without requiring major 
resources~(e.g.~\cite{Shibata:2004nv,Duez:2006qe,Kiuchi:2008ss}).
In contrast, studies of the most general flows and instabilities require 3D simulations and are
the most expensive realistically accessible scenarios being currently considered. Future efforts, including
radiation transport mechanisms, will move beyond 3D scenarios and require efficient use of petaflop (and beyond)
resources (e.g. ~\cite{petscale}).

Another important distinction in these studies is whether the modeled stars are
rigidly rotating or allow for differential rotation. Rigidly rotating stars
support more mass than non-rotating ones (so-called TOV stars), but generally
do not support the largest masses achieved with differential rotation 
nor do they demonstrate some of the more interesting
instabilities such as the bar mode instability~\cite{Zink:2006qa}.
Instead, uniformly rotating stars tend to demonstrate one of two behaviors;
stability or instability to collapse to a black hole. Previous studies
suggest that significant disks do not form from such
collapse~\cite{Shibata:2003iw,Duez:2004uh,Baiotti:2004wn}. Recall that
stars rotating differentially are expected to settle
into rigidly rotating configurations on short time scales, and hence a normal
neutron star observed today is generally expected to be rigidly rotating.

Achieving ever more realism, models of neutron stars have also begun
to consider 
stars with magnetic field~\cite{Stephens:2008hu,Duez:2006qe,Shibata:2005mz,Shibata:2006hr}.
Magnetic fields provide, in particular, an effective pressure which generally supports greater mass~\cite{Bocquet:1995je} as well as an efficient way to transport angular momentum. In
differentially rotating stars, even small magnetic fields can be amplified
by the magnetorotational instability. Generally these studies begin with a nonmagnetized
neutron star to which a small, seed magnetic field is added. However, in~\cite{Giacomazzo:2007ti},
fully consistent, magnetized stars are evolved, although only nonrotating results are presented.
Another approach is to study the modes through a perturbation approach~\cite{Sotani:2006yg,Sotani:2006at} and analyze the growth of these modes with respect to
a given stationary solution. The dynamical behavior of magnetized stars is important also for their
role in explaining strong electromagnetic emissions. Indeed, 
isolated stars with strong magnetic fields, so called {\em magnetars}, are suspected 
to be the engines powering anomalous X-ray pulsars~(AXPs) and soft gamma ray repeaters~(SGRs)~\cite{Mereghetti:2008je}. At least $10\%$ of all neutron
stars~\cite{Beloborodov:2006qh} are born as magnetars.
In the context of single stars, it is thus interesting to examine if strong magnetic fields may deform stars away from axisymmetry making them strong producers of gravitational waves~\cite{Haskell:2007bh}.

Here we present results with uniformly rotating neutron stars which possess a fully
consistent magnetic dipole moment. That is to say, the initial data used here represents
a stationary state of the full GR+MHD equations. These evolutions are computed with  
a general relativity code employing the generalized harmonic scheme (allowing for 
black hole excision). Further,
no symmetry assumptions are made. High resolutions are achieved using a distributed adaptive
mesh refinement infrastructure. These evolutions demonstrate that our code
can evolve a stable rotating star for many periods accurately. Similarly, unstable stars
evolve to black holes with no evidence of any significant disk forming. Finally, we give
evidence that unstable, rotating, magnetized stars represent minimally unstable solutions
which could serve as Type I critical solutions.

In Section~\ref{sec:eom} we provide details about the formulation of
the equations. In Section~\ref{sec:implementation} we discuss aspects of our
numerical implementation, and describe the diagnostic quantities evaluated
in Section~\ref{sec:diags}. In Section~\ref{sec:id} we discuss the
initial data we use. We present our results in Section~\ref{sec:results},
and conclude in Section~\ref{sec:conclusions}.

%
%
\section{Formulation and equations of motion}
\label{sec:eom}

Neutron stars can be modeled by relativistic fluids (possibly with the
inclusion of magnetic fields) under the action of strong gravitational fields~\cite{Andersson:2006nr}.
These systems are governed both by the Einstein equations for the
geometry and by the relativistic equations of magnetohydrodynamics for the matter.
We write both systems as first order hyperbolic equations. This form of 
the equations is convenient in order to take advantage of several rigorous numerical 
techniques devised for such systems to ensure, at the linear level, stability of
the implementation. More information regarding
the motivation for this approach can be found in~\cite{Palenzuela:2006wp,Neilsen2005,Anderson:2006ay}.
By way of notation, we use letters from the beginning of the alphabet
($a$, $b$, $c$) for spacetime indices, while letters from the middle
of the alphabet ($i$, $j$, $k$) range over spatial components.
We adopt geometric units where $c=G=1$. However, as discussed in Section~\ref{sec:implementation},
when appropriate, we rescale the value of $G$ to achieve improved accuracy
in the conservative to primitive variable conversion stage.

\subsection{The Einstein equations}
\label{sec:eom:ee} 
The Einstein equations can be written as a system of ten nonlinear partial
differential equations for the spacetime metric $g_{ab}$. The harmonic formulation
of the Einstein equations exploits the fact that the coordinates $x^a$ can be chosen
satisfying the generalized harmonic condition~\cite{Frie85,Garfinkle} 
\begin{equation}
\label{eq:harmonic}
\nabla^c \nabla_c x^a = - \Gamma_a = H^a(t,x^i) \label{harmonic1} \, ,
\end{equation}
where $\Gamma^a \equiv g^{bc} \Gamma_{abc}$ are the contracted Christoffel symbols.
The arbitrary source functions $H^a(t,x^i)$ determine the coordinate freedom
of Einstein equations. The original harmonic coordinates correspond to the case
$H^a(t,x^i)=0$, which is the choice here. The Einstein equations can be expressed
in their generalized harmonic form~\cite{Frie85}, in particular
\begin{eqnarray}\label{dedonder1}
 && g^{cd} \partial_{cd} ~g_{ab}
 + \partial_{a} H_{b} + \partial_{b} H_{a}  =
 - 16~\pi~\left(T_{ab} - \frac{T}{2}~g_{ab}\right) \nonumber \\
 && + 2~\Gamma_{cab}H^c
  + 2~g^{cd}g^{ef}\bigg(\partial_e g_{ac}~ \partial_f g_{bd}
 - \Gamma_{ace}~\Gamma_{bdf}\bigg).
\end{eqnarray}
The matter is coupled to the geometry by means of the stress energy
tensor $T_{ab}$ and its trace $T\equiv g^{ab} T_{ab}$, which will be dictated
by the particular model of magnetized fluid under consideration, detailed 
in the next subsection.

The spacetime can be foliated into hypersurfaces of constant coordinate
time $x^0\equiv t = {\rm const}$. 
On these spacelike hypersurfaces, one defines a spatial 3-metric
$h_{ij} = g_{ij}$. A vector normal to the hypersurfaces is given by
$n_a \equiv - \nabla_a t / ||\nabla_a t ||$, and coordinates defined on 
neighboring hypersurfaces can be related through the lapse, $\alpha$, 
and shift vector, $\beta^i$.  With these definitions, the spacetime
differential element can then be written as
\begin{eqnarray}
\rmd s^2 &=& g_{ab}\, \rmd x^a \rmd x^b \nonumber \\
         &=&-\alpha^2 \, \rmd t^2 
            + h_{ij}\left(\rmd x^i + \beta^i\, \rmd t\right)
                    \left(\rmd x^j + \beta^j\, \rmd t\right).
\end{eqnarray}
Indices on spacetime quantities are raised and lowered with the 4-metric,
$g_{ab}$, and its inverse, while the 3-metric $h_{ij}$ and its inverse
are used to raise and lower indices on spatial quantities.

We adopt a first order
reduction of the second order differential equations represented in
Eqs.~(\ref{dedonder1}).
This reduction can be achieved by introducing new independent variables related
to the time and space derivatives of the fields
\begin{equation}
 \quad Q_{ab} \equiv - n^c \, \partial_c g_{ab} \, ,
\quad  D_{iab} \equiv \partial_i g_{ab} \, .
\label{eq:firstorder}
\end{equation}
Within these definitions we can write our evolution equations in our GH
formalism in the following way \cite{Lindblom:2005qh}
\begin{eqnarray}
  \label{EE_geq}
  \partial_t g_{ab} &=& \beta^k~D_{kab} - \alpha~Q_{ab}, \\
  \partial_t Q_{ab} &=& \beta^k~\partial_k Q_{ab}
  - \alpha h^{ij} \partial_i D_{jab} \nonumber \\ 
  &-& \alpha~ \partial_a H_b - \alpha~ \partial_b H_a +
  2~\alpha~ \Gamma_{cab}~ H^c \nonumber \\
  &+& 2\, \alpha\, g^{cd}~(h^{ij} D_{ica} D_{jdb} - Q_{ca} Q_{db}
                   - g^{ef} \Gamma_{ace} \Gamma_{bdf}) \nonumber \\
  &-& \frac{\alpha}{2} n^c n^d Q_{cd} Q_{ab}
  - \alpha~h^{ij} D_{iab} Q_{jc} n^c \nonumber \\ 
  &-& 8 \pi \, \alpha(2T_{ab} - g_{ab} T) \nonumber \\
  &-& 2 \sigma_0 \, \alpha \, [n_a Z_b + n_b Z_a - g_{a b} n^c Z_c ]  \nonumber  \\
  &+& \sigma_1 \, \beta^i ( D_{iab} -  \partial_i g_{ab} ),  \\
   \label{EE_Deq}
   \partial_t D_{iab} &=& \beta^k \partial_k D_{iab}
  - \alpha~\partial_i Q_{ab} \nonumber \\ 
   &+& \frac{\alpha}{2} n^c n^d D_{icd} Q_{ab}
  + \alpha~h^{jk} n^c D_{ijc} D_{kab} \nonumber \\ 
   &-& \sigma_1 \, \alpha \, ( D_{iab} - \partial_i g_{ab} ) .
  \label{EE_geqend}
\end{eqnarray}
This GH formulation includes a number of constraints that must be satisfied 
for consistency. On one hand, there are two sets of first order constraints,
obtained from Eqs.~(\ref{eq:firstorder}) and defined as
\begin{eqnarray}\label{firstorder_constraint}
  {\cal C}_{iab} &\equiv & \partial_i g_{ab} - D_{iab} = 0 ~,~~\nonumber \\
  {\cal C}_{ijab} &\equiv & \partial_i D_{jab} - \partial_j D_{iab} = 0 ~,
\end{eqnarray}
which were introduced when performing the reduction to first order
~\cite{Lindblom:2005qh,Palenzuela:2006wp}. On the other hand, there
are the Hamiltonian and momentum constraints, that in the Generalized 
Harmonic formulation show up in terms of a four-vector $Z_a$, which 
is defined as
\begin{equation}\label{harmonicZ}
    2 Z^a \equiv -\Gamma_a - H^a(t,x^i) \, .
\end{equation}
It can be shown that the Hamiltonian and momentum constraints are satisfied if
$Z^a=\partial_t Z^a=0$ \cite{BLPZ04}. In order to dynamically control the
violation of the constraints, we have included certain terms proportional
to these constraints (\ref{firstorder_constraint})--(\ref{harmonicZ}). 
These additional terms depend on free parameters $\sigma_0$ and $\sigma_1$,
allowing one to dynamically damp constraint---including the 
Hamiltonian, momentum, and first-order constraints 
(\ref{firstorder_constraint})---violating
modes on a time scale proportional to $-\sigma_i$ (\cite{Gundlach,Lindblom:2005qh}). 

We evolve the gravitational field equations shown in Eqs.~(\ref{EE_geq}-\ref{EE_geqend}).
These equations rely on the computation of
the 4-dimensional
Christoffel symbols from the metric $g_{ab}$
\begin{equation}
  \Gamma_{abc} = \frac{1}{2} \, (D_{bca} + D_{cba} - D_{abc}) ~.
\end{equation}
While we evolve the $D_{iab}$ functions, the
set $D_{0ab}$ are not evolved, but are calculated from evolved quantities as
\begin{equation}
  D_{0ab} = -\alpha Q_{ab} + \beta^k D_{kab} \, .
\end{equation}

This description suffices to explain the gravitational evolution, and
the following section describes the evolution of the matter. However, we note here
that the MHD equations are written
in the standard 3+1 decomposition of spacetime and thus require the
spatial metric $h_{ij}$,
the lapse $\alpha$, shift $\beta^i$, and ADM extrinsic curvature, $K_{ij}$. 
These quantities
can be written in terms of our evolved fields using
\begin{eqnarray}
  h_{ij} &=& g_{ij}~,~~\alpha = \sqrt{-1/g^{00}}~,
        ~~\beta^i = \gamma^{ij} g_{0j}~,\nonumber \\
  K_{ij} &=& \frac{1}{2} Q_{ij}+\frac{1}{\alpha}(D_{(ij)0}-\beta^k D_{(ij)k}).
\end{eqnarray}
Conversely, the Hamiltonian and momentum constraints are usually
written in terms of spatial derivatives of the metric $D_{kij}$ and
the extrinsic curvature $K_{ij}$.  In fact, we use these 3+1
quantities (and similar expressions for their derivatives) to
calculate the residuals of the Hamiltonian and momentum constraints
expressed in their standard form.

%
%
\subsection{MHD equations}

We now briefly introduce the perfect fluid equations.  
Additional information can be found in our previous 
work~\cite{Neilsen2005,Anderson:2006ay}
as well as in topical review articles~\cite{Marti:1999wi,Font:2000pp}.

The stress-energy tensor for the perfect fluid in the presence of 
a Maxwell field is given by
\begin{eqnarray}\label{stressenergy}
T_{ab} &=& \left[ \rho_o(1+\epsilon) + P \right] u_a u_b + P g_{ab} 
\nonumber \\
      &+& {F_a}^c F_{bc} - \frac{1}{4} g_{ab} F^{cd} F_{cd} ~.
\end{eqnarray}
The fluid is described by rest mass density $\rho_o$, the specific
internal energy density $\epsilon$, the isotropic pressure $P$ and
the four velocity of the fluid $u^a$. With these quantities we can
construct the enthalpy
\begin{equation}
h_e = \rho_o + \rho_o\epsilon + P,
\label{eq:enthalpy}
\end{equation}
and construct the standard  spatial coordinate velocity of the fluid
$v^i$ as
\begin{equation}
W\equiv -n^a u_a, \qquad v^i \equiv \frac{1}{W}\,h^i{}_j u^j,
\end{equation}
where $W$ is the Lorentz factor between the fluid frame and the fiducial ADM
observers.

The Maxwell tensor $F_{ab}$ can be written as
\begin{equation}\label{F_em1a}
  F^{ab} = n^{a} E^{b} - n^{b} E^{a} + \epsilon^{abcd}~B_{c}~n_{d},
\end{equation}
where $E^{a}$ and $B^{a}$ are the electric and magnetic fields measured by 
a ``normal'' observer $n^{a}$. Consequently, both fields are purely spatial,
i.e., $E^{a} n_{a} = B^{a} n_{a} = 0$.

The evolution of the magnetized fluid is described by different sets of conservation laws. The magnetic field, in the ideal MHD limit, follows the Maxwell equation
\begin{eqnarray}
  \nabla_{\mu} (^*F^{\mu \nu} + g^{\mu \nu} \Psi) &=& \kappa n^{\nu} \Psi \, ,
\label{Maxwell_ext_eqs2b} 
\end{eqnarray}
where $^*F^{ab} \equiv \epsilon^{abcd} F_{cd}/2$ is the dual of the Maxwell
tensor and we have introduced a real scalar field $\Psi$ to control the divergence constraint.
This technique is known as divergence cleaning~\cite{dedner2002} and allows for a convenient
way to control the constraint violation by inducing a damped wave equation for the scalar field $\Psi$.
The other Maxwell equation, in the ideal MHD limit, only
gives the definition for the current density, since the electric field
is given in terms of the velocity of the fluid and the magnetic field,
that is,
\begin{eqnarray}
    E_i = -\epsilon_{ijk} v^k B^k \, .
\end{eqnarray}

Conservation of the stress energy tensor in Eq.~(\ref{stressenergy}),
\begin{eqnarray}
  \nabla_{a} T^{ab} = 0,
\end{eqnarray}
provides 4 evolution equations for the fluid variables, namely the
velocity and the internal energy. Conservation of the baryon number
\begin{eqnarray}
  \nabla_{a} (\rho_o u^a) = 0
\end{eqnarray}
leads to the evolution equation of the rest mass density $\rho_o$. Closure
of the equations is achieved by introducing an equation of state~(EOS)
relating the pressure with the other thermodynamical quantities of the fluid, $P=P(\rho_o,\epsilon)$. 

High resolution shock capturing schemes (HRSC) are robust numerical methods for
compressible fluid dynamics.  These methods, based on Godunov's seminal
work~\cite{Godunov}, are fundamentally based on expressing the fluid equations as
integral conservation laws.  To this end, we introduce {\em conservative}
variables ${\bbq} = (D, S_i, \tau, B^i)^{\rm T}$, where
\begin{eqnarray}
D &=& W \rho_o,\\
S_i &=& (h_e W^2 + B^2) v_i - (B^j v_j) B_i,\\
\tau &=& h_e W^2 + B^2 - P \nonumber\\
     & & \quad - \frac{1}{2}\left[ (B^i v_i)^2 + \frac{B^2}{W^2} \right] - D.
\end{eqnarray}
and $B^i$ is both a primitive and conservative variable.
In an asymptotically flat spacetime these quantities are conserved, 
and are related to the total energy, momentum, and, in the non-relativistic limit, 
the kinetic energy, respectively.  
The quantities ${\bbw} = (\rho_o , v^i , P, B^i)^{\rm T}$ are called
the {\em primitive} variables in contrast to the conservative variables.
The fluid state can be specified using either set of variables, and both sets
are required to write the MHD evolution equations.
Anticipating the form of these equations,
we also introduce the densitized conserved variables
\begin{equation}
\tilde D = \sqrt{h}\, D, \quad 
\tilde S_i = \sqrt{h}\, S_i, \quad
\tilde \tau = \sqrt{h}\, \tau, \quad
\tilde B^i = \sqrt{h}\, B^i, \quad \label{eq:defineconserve}
\end{equation}
where $h=\det(h_{ij})$. 
The fluid equations can now be written in balance law form
\begin{equation}
\partial_t\tilde\bbq  + \partial_k\bbf\,^k(\tilde\bbq) = \bbs(\tilde\bbq),
\label{eq:balance}
\end{equation}
where $\bbf\,^k$ are flux functions, and $\bbs$
are source terms.  The fluid equations in this form are
\begin{eqnarray}
&&\partial_t \tilde D + \partial_i \left[ \alpha\,\tilde D
\left( v^i - {\beta^i \over \alpha} \right) \right] = 0,\label{eq:ev_D} \\
&& \partial_t \tilde S_j + \partial_i \left[ \alpha \left(
\tilde S_j \left( v^i - {\beta^i \over \alpha} \right) + \sqrt{h}\,P \, h^i{}_j
   \right)\right]\nonumber\\
&&\qquad  = \alpha \,  {^{3}{\Gamma}}^i{}_{jk} \, \left( \tilde S_i v^k 
         + \sqrt{h}\,P h_i{}^k \right)  
   + \tilde S_a\partial_j\beta^a\Bigr.\nonumber\\
&&\qquad\qquad\qquad
  - \partial_j \alpha \, (\tilde\tau + \tilde D) \nonumber \\
&& \qquad\qquad\qquad - \zeta \alpha (\tilde B_i W^{-2} + v_i v_j \tilde B^j )  \partial_k \tilde B^k , \\
 &&\partial_t \tilde\tau
        + \partial_i \left[ \alpha\left(\tilde S^i - \frac{\beta^i}{\alpha} \, 
                     \tilde\tau - v^i \tilde D \right) \right] \nonumber\\
&& \qquad= \alpha \, 
\left[ K_{ij} \tilde S^i v^j + \sqrt{h}\, K P - \frac{1}{\alpha} 
\, \tilde S^a \partial_a \alpha \right],\nonumber \\
&& \qquad\qquad\qquad  -\zeta \alpha v_j \tilde B^j \partial_k \tilde B^k \\
&& \partial_t \tilde B^b
+ \partial_i \left[ \tilde B^b \left( v^i - \frac{\beta^i}{\alpha} \right)
                   -\tilde B^i \left( v^b - \frac{\beta^b}{\alpha} \right)
                   \right] \nonumber\\
&& \qquad= - \alpha \sqrt h h^{bi} \partial_i \Psi 
- \zeta \alpha v^i \partial_j \tilde B^j \label{eq:ev_B} \\ 
&&\partial_t \Psi = - c_r \alpha \Psi - c_h \frac{\alpha}{\sqrt h}
                 \partial_i \tilde B^i +
		 (\beta^i - \alpha v^i) \partial_i \Psi  \, .       		   
\label{eq:ev_Psi}
\end{eqnarray}
where $c_r = \kappa$ and we have allowed for different speeds than light
by introducing the parameter $c_h$. Except in the tests, we will use the original
prescription (\ref{Maxwell_ext_eqs2b}) with $c_h=1$.
Here ${^{3}{\Gamma}}^i{}_{jk}$ is the Christoffel symbol associated with the
3-metric $h_{ij}$, and $K$ is the trace of the extrinsic curvature, 
$K = K^i{}_i$. Notice that the aforementioned system is an extended
version  of the one employed in our earlier works~\cite{Neilsen2005,Anderson:2006ay,Anderson:2008zp}
Here we have added additional terms toggled by  the 
parameter $\zeta$ which allow for considering an extension of the
``eight-wave'' formulation which, in the absence of the cleaning field $\Psi$, ensures
the strong hyperbolicity of the system~\cite{dedner2002} --at the cost of introducing
derivative terms in the sources--. Furthermore, 
by setting $\zeta=1$, the propagation speeds of two constraint violating modes 
become non-vanishing and hence violations are dragged along by the fluid's velocity.
This is numerically convenient
as possible violations will propagate off the grid.

Finally, we close the system of fluid equations with an equation of state (EOS).
We choose the ideal fluid EOS
\begin{equation}
P = (\Gamma-1)\, \rho_o\epsilon,
\end{equation}
where $\Gamma$ is the constant adiabatic exponent.  Nuclear matter in 
neutron stars
is relatively stiff, and we set $\Gamma=2$ in this work.
When the fluid flow is adiabatic, this EOS reduces to the well known
polytropic EOS 
\begin{equation}
P=\kappa\rho_o{}^\Gamma,
\label{eq:polyEOS}
\end{equation}
where $\kappa$ is a dimensional constant.   We use the polytropic EOS
only for setting initial data. 

The set of fluid equations, Eqs.~(\ref{eq:ev_D}-\ref{eq:ev_B}), are used
to evolve the conservative variables. However, these equations also contain the
primitive variables which necessitates a step in the evolution scheme which solves for 
the primitive variables in terms of the conservative ones. 
Given the primitive variables, the conservative variables are easily
calculated from the algebraic expressions in Eqs.~(\ref{eq:defineconserve}).  Calculating the
primitive variables from the conservative variables, however, is more
delicate, as it requires the solution of a transcendental equation. To this end we first
define the quantity $x\equiv h_e W^2$.  We then write $S^2=S^i S_i$ in terms
of $x$ and solve for $W^2$, obtaining
\begin{equation}
W^2 = \frac{x^2(x + B^2)^2}{x^2(x+B^2)^2 - x^2S^2 - (2x+B^2)(S_iB^i)^2}.
\label{eq:w2_x}
\end{equation}
Using the definition of $\tau$, we define a function that is identically zero
\begin{equation}
f(x) = x - P - \frac{1}{2}\left[\frac{(S_iB^i)^2}{x^2} + \frac{B^2}{W^2}\right]
+ B^2 - \tau - D = 0.
\end{equation}
If the enthalpy can be expressed as a simple function of the pressure, 
as can be done for the ideal fluid and a generalized EOS with a cold
nuclear component, then
we can express the pressure as a function $x$.
For the ideal fluid EOS used here, the enthalpy equation
\begin{equation}
h_e = \frac{x}{W^2} = \rho_0 + \rho_0\epsilon + P
\end{equation}
can be solved for $P$ by substituting in $\rho_0=D/W$ and the EOS to obtain
\begin{equation}
P = \frac{\Gamma-1}{\Gamma}\left(\frac{x}{W^2} - \frac{D}{W}\right).
\end{equation}
Combining these equations, $f$ is a function of a single unknown $x$.  This
equation can be solved numerically using the Newton-Raphson method.  
It is useful to note that $x$ has a
minimum physical value, which is found by requiring in Eq.~(\ref{eq:w2_x})
that $W^2\ge 1$.

\begin{figure}
\begin{center}
\epsfig{file=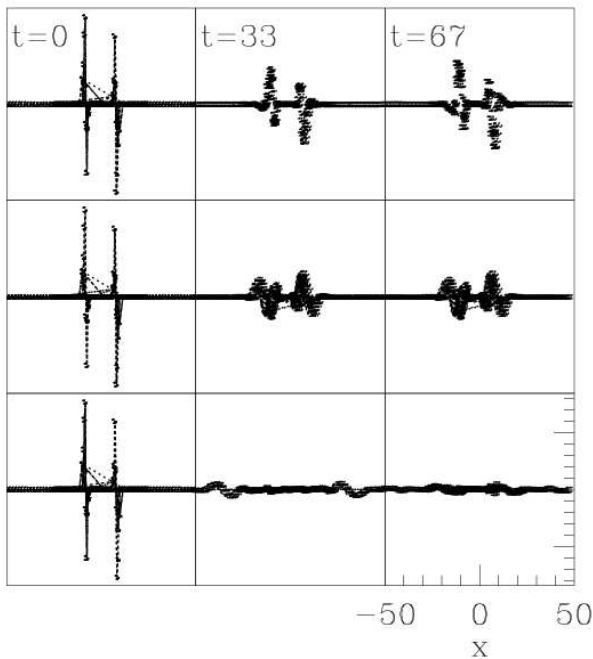,height=9.0cm}
\caption{Demonstration of the effectiveness of divergence cleaning at handling deviations from a divergenceless
    magnetic field. A calculation of the divergence of the magnetic field, $x^2\left[\vec{\nabla}\cdot \vec{B}\right]$,
     along the $x$-axis
    for three times is shown for each of
    three cases: (top) No divergence cleaning; (middle) Cleaning with $c_{\rm h}=0.1$ and $c_{\rm r}=0.01$; and
    (bottom) Cleaning with $c_{\rm h}=1.0$ and $c_{\rm r}=0.1$. These runs were otherwise identical for a
    magnetized, rotating star of coordinate equatorial radius of $10$ with spin along the $z$-axis 
    with three levels of refinement
    (at $\pm 50$, $\pm 25$, and $\pm 12.5$) and a perturbation to the magnetic field to
    introduce an explicit deviation from divergenceless.
    Concentrating on the MHD equations, the metric terms were frozen at their initial values,
    e.g.\ the Cowling approximation.
     The bottom row shows clear wavelike behavior as it
    ``cleans'' the divergence.
     } \label{fig:divb}
\end{center}
\end{figure}

\section{Implementation Details}
\label{sec:implementation}

The code is constructed within the \had\ computational infrastructure
which provides distributed adaptive mesh refinement~(AMR).
The AMR follows in the style of Berger \& Oliger~\cite{berger}, but
uses the tapering condition for AMR boundaries instead of temporal
interpolation~\cite{Lehner:2005vc}. 
The combined set of geometric equations
and fluid equations, Eqs.~(\ref{EE_geq}-\ref{EE_geqend}) and
Eqs.~(\ref{eq:ev_D}-\ref{eq:ev_Psi}) respectively, is discretized using
the method of lines.  The geometric equations are discretized using
operators that satisfy a summation by parts property~\cite{Strand1994a,Calabrese:2003a}.  
The fluid equations
are discretized using the HLLE method~\cite{Harten83}.  The semi-discrete equations
are solved using a third order accurate, total variation diminishing~(TVD) 
Runge-Kutta solver~\cite{ShuOsherII}.

The fluid equations diverge as the density goes to zero,
and, as is standard practice, we disallow
any true vacuum by setting such regions to a floor or atmosphere value. 
The floor is applied after each fluid update as
\begin{equation}
D \leftarrow \min(D, \delta), \quad \tau \leftarrow \min(\tau, \delta), 
\end{equation}
where $\delta$ is chosen to
be many orders of magnitude smaller than the maximum densities and
pressures in the initial data. The comparison of otherwise identical
runs but with different floor values suggest that the use of an
atmosphere generally does not affect accuracy.

We have, however, found certain issues with precision occurring within
the primitive solver. Typical maximum values of the density are about $10^{-2}$
in geometric units,
with a floor value of $\delta=10^{-8}$. We have found it useful therefore to
scale Newton's constant $G$ such that the fluid densities and pressures
are close to order unity.  Thus, rather than using the typical choice
of $G=1$, we might use $G=1/1000$.
As $G$ affects only the coupling of the fluid to the geometry,
the evolution of the geometric equations is not affected by
this scaling.
Empirically, we find that
scaling G allows the primitive variables to be more easily recovered
in low density regions. 
This improvement appears to be related to finite precision effects
in the primitive solver.
The scaling decreases the effective floor value
while avoiding the problems associated with having a true vacuum.

The Maxwell equations require that the magnetic field be divergenceless. 
This is the so-called ``no monopole''
constraint. A variety of schemes exist with the goal of controlling
the growth of the divergence. We choose a strategy that ensures flexibility
and robustness when dealing with multiple grid structure (as in AMR) and 
allows, in principle, for a clean boundary treatment~\cite{Cecere:2008sj}.
To this end we have implemented hyperbolic divergence cleaning
as described in~\cite{dedner2002} (also see~\cite{toth}). We thus introduce
a scalar $\Psi(x,y,z,t)$ which  is sourced by the negative of the
divergence of the magnetic field as shown in Eq.~(\ref{eq:ev_Psi}). 
As described in~\cite{dedner2002}, this
scheme implies that the divergence obeys a damped wave equation so that
constraint violations propagate off the grid and their value is reduced.

For initial data generated with \magstar, the divergence is around
machine precision, and so to test the implementation of divergence
cleaning, we introduce a perturbation to the magnetic field in order
to produce a significant amount of divergence to the magnetic field.
In particular, this perturbation takes the form of a spherical,
Gaussian shell of  radius $r_o$, width $\delta$, and amplitude $A$
added to each component of $\vec{B}$. We expect the divergence
cleaning to propagate this perturbation 
as a damped wave, and we therefore plot the scaled quantity
$x^2 \left[ \vec{\nabla}\cdot \vec{B} \right]$ in Fig.~\ref{fig:divb}.
As can be seen by comparing the cases of no cleaning ({\em top}) 
and with cleaning ({\em bottom}), the divergence
propagates with damping through  the refinement 
boundaries across the grid.

As typical with codes dealing with linearly degenerate hyperbolic
systems, like those those employed in numerical relativity, a dissipation
operator is applied to the metric variables. This operation uses
a high-order derivative to serve as a low-pass filter and does not
affect the accuracy of the simulation. We have found it useful for
keeping things smooth. Although this operation is not applied to
the fluid variables, we have found it quite important in keeping
the magnetic field components smooth. The magnetic
field evolution is coupled tightly with that of the velocity, and
any nonsmoothness which appears in the velocity can easily affect
the smoothness of the magnetic field. The addition of dissipation
to the magnetic field  and divergence cleaning field
helps control the behavior of the magnetic field.

At the boundaries of the domain, simple outflow boundary conditions are
applied to the fluid variables. This is accomplished by copying the values
of the conservative variables near the boundary outward.
Most of the gravitational variables are treated using Sommerfeld-like
boundary conditions of the form ~\cite{RLS07}
\begin{equation}
   \left(\partial_t +\partial_r + \frac{1}{r}\right)\left(g_{ab} - \eta_{ab}\right) = 0 ~~,
\label{Sommerfeld1}
\end{equation}
where $\eta_{ab}$ is just the Minkowski metric. The rest, which
are not so crucial, are set either by maximally dissipative
\cite{Palenzuela:2006wp} or constraint preserving boundary conditions \cite{Lindblom:2005qh}.\\

A relevant issue when considering the collapse of stars is the formation
of a black hole. To deal with such situations, we adopt black hole excision 
where we dynamically monitor for the appearance of trapped surfaces (which
lie inside an event horizon if cosmic censorship holds) and excise cubical region(s)
from the computational domain. As discussed in~\cite{Tiglio:2003xm,Calabrese:2003vx}, this excision
introduces inner boundaries
which are of ``outflow'' type and so no boundary condition is required there.
However, for a more robust handling of the fluid, we also allow for a modification of the
fluid equations inside the trapped surface~\cite{Megevand:2009yx}.
The MHD equations are written in balance law form
\begin{equation}
\dot U + F(U)' =  S,
\end{equation}
which we modify to include a damping term near the black hole
\begin{equation}
\dot U + F(U)' =S -f(r) {(\Delta x)}^p (U-U_0).
\end{equation}
Here the function $f(r)$ decreases smoothly with $r$, 
 from a given value at the excision region to
zero at the outermost trapped surface~(OTS) found,
and is zero for $r\geq r_{\rm OTS}$,
so that the
exterior of the BH is causally disconnected from the effect of this extra
term. $U_0$ is set to zero or to the value of the atmosphere if the
corresponding field has one.
The coefficient ${(\Delta x)}^p$ ensures that the damping term converges
to zero as the grid spacing $\Delta x$ is reduced. As long as one chooses $p$
greater than or equal to the order of convergence of the code, this term
will not modify the convergence rate.
We typically adopt a value of $p=4$.

Finally, gravitational radiation is calculated via the evaluation of the Newman-Penrose scalar
$\Psi_4$ which is computed by contracting the Weyl tensor, $C_{abcd}$, 
with a suitably defined null tetrad $\{\ell,n,m,\bar m\}$
\begin{eqnarray}
 \Psi_4 = C_{abcd} n^a \bar m^b n^c \bar m^d
\end{eqnarray}
extracted at spherical surfaces $\Sigma_i$ located in the wave-zone,
far from the sources. We also consider possible corrections required to deal 
with gauge ambiguities, as discussed in \cite{Lehner:2007ip}. We refer the reader to
that paper for details on the adopted tetrad and required corrections.

\section{Diagnostic Quantities}
\label{sec:diags}

The initial configuration for the stars is axisymmetric, and we therefore
want to be able to measure any change to this structure. For this purpose,
we monitor certain
distortion parameters~\cite{Baiotti:2006wn} defined as
\beq
\eta_+      & = & \frac{ I^{xx} - I^{yy} }
                       { I^{xx} + I^{yy} }  \\
\eta_\times & = & \frac{ 2 I^{xy} }
                       { I^{xx} + I^{yy} }  \\
\eta        & = & \sqrt{\eta^2_+ + \eta^2_\times } 
\eeq
in terms of the moment of inertia tensor
\be
I^{jk} \equiv \int D\, x^j x^k~d^3x.
\ee
These parameters are computed with respect to the conservative variable $D$.

It is also standard practice to display the maximum of the (primitive) rest mass density, and
we compute the fractional change in time as
\be
\Delta \rho \equiv \frac{ {\rm max}\left|\rho_0(t)\right|-{\rm max}\left|\rho_0(0)\right| }
                        { {\rm max}\left|\rho_0(0)\right| }.
\ee
Generally, for stable stars one sees this quantity oscillate from inherent numerical
perturbations about a stable solution. For unstable solutions, one expects this quantity
to change in a significant way.
Similarly, we compute the relative change in baryon mass as a function of time as
\be
\Delta M_{\rm baryon} \equiv \frac{M_{\rm baryon}(t)-M_{\rm baryon}(0)   }
                         { M_{\rm baryon}(0) }
\ee
 where
\be
M_{\rm baryon} \equiv \int D~dV.
\ee
This mass is related to the expected baryon number and should be strictly conserved
as long as mass is not leaving the computational domain. Many HRSC schemes explicitly
conserve this quantity, but here it is not {\em a priori} conserved. Our use of an
atmosphere entails adding a small amount of mass in regions that would otherwise
become evacuated.  Another reason  is that we are using a finite difference based
AMR which does not accommodate as readily a strictly conservative treatment as a finite
volume method would~\cite{colella}.  
Finally, the presence of 
source terms in the evolution equations for the other conservative 
variables also breaks perfect conservation of these other quantities.

We also compute the  angular momentum of the fluid. Since our
stars rotate about the $z$-axis, we need only compute a single such quantity
\be
J_z = \int \left( x n^j T_{jy} -y n^j T_{jx} \right) \sqrt{h}~d^3x.
\ee
The use of Cartesian coordinates is typically avoided when one expects
angular momentum to be conserved since cylindrical coordinates are
better adapted to the respective Killing vector.
However, because this
project is part of a more general effort to model a variety of systems with
different natural coordinates and no symmetries, we simply monitor the extent to which this is
conserved by computing a fractional change as
\be
\Delta J_z \equiv \frac{J_z(t)-J_z(0)   }
                         { J_z(0) }.
\ee

Finally, we also monitor the extent to which our numerical solution
satisfies the Einstein equations as expressed in Eq.~(\ref{eq:harmonic}).
That is,  we compute the so-called residual of these equations, expressed
as the four-vector $Z^a$ as in Eq.~(\ref{harmonicZ}).  In particular, 
we monitor the norm of this vector
\be
|\vec{Z}| \equiv ||Z^a||_2.
\ee
Analytical solutions to Einstein equations must have a vanishing residual and thus we monitor
this residual for solutions obtained at various resolutions to check for
convergence to a consistent solution.

\section{Initial Data}
\label{sec:id}

\begin{figure}
\begin{center}
\epsfig{file=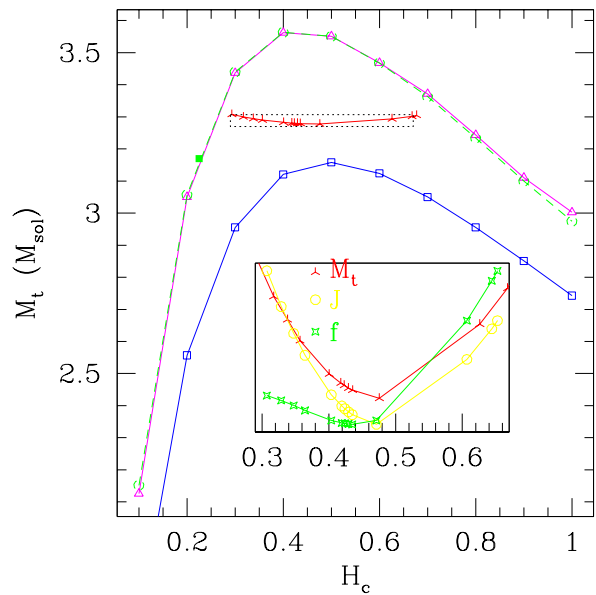,height=9.0cm}
\caption{Diagram of solution space of stars generated with \magstar\  (using its units).
     The total gravitational mass is plotted versus the central enthalpy.
     The upper curve (purple, open triangles) represents stars with no magnetic field at
     the mass shredding limit. Very close to this are shown (green, open circles)
     stars rotating at the same frequencies but with a radial magnetic field at the pole 
     of $1000~\gigatesla=10^{16} G$. The masses of these stars are barely changed with respect to their
     nonmagnetic counterparts.
     The lower curve (blue, open squares) 
     represents the static limit with no rotation. A sequence of 
     constant baryon mass stars ($M_B=3.6 M_{\rm sol}$) is shown (red stars).
     This sequence is also shown in the inset along with their angular
     momentum (yellow open circles) and frequency of rotation (green stars).
     The location of the initial data used in
     the initial data convergence test of Fig.~\ref{fig:idconv}
     and
     the long term, stability test of Fig.~\ref{fig:stable}
     is also shown (green, solid square) although that solution
     strictly does not belong here because it has non-vanishing magnetic
     field.
     } \label{fig:solutionspace}
\end{center}
\end{figure}

\begin{figure}
\begin{center}
\epsfig{file=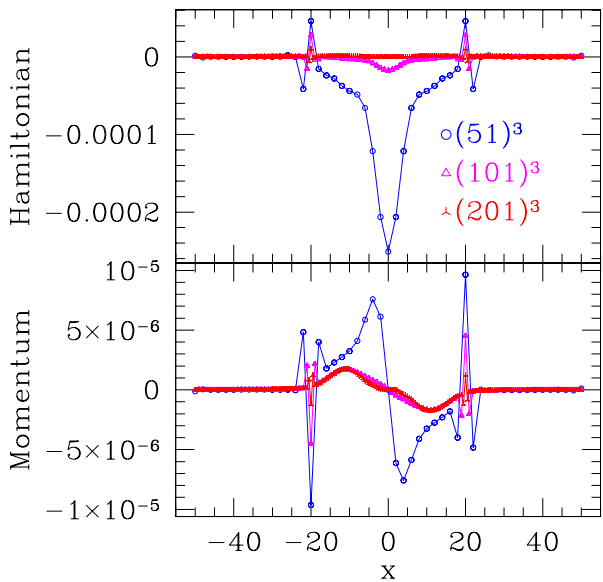,height=9.0cm}
\caption{The residuals of two constraint equations, namely the Hamiltonian
     constraint and the $y$-component of the momentum constraint.
     A slice along the $x$-axis 
     of the computed residuals for various unigrid resolutions.
     {\bf Top:} That the Hamiltonian constraint residual converges to zero with
     increasing resolution is taken as evidence that the initial
     data is properly constructed and read into the code. {\bf Bottom:} That the
     momentum constraint residual improves with the first increase
     in resolution is similar evidence. However,
     the next increase in resolution  fails to bring
     down this residual. We suspect that this remaining error is associated 
     not with truncation error but instead with
     inherent errors in the numerical transformation and interpolation
     from \lorene's spherical basis  to our Cartesian one.
     Note the spikes that arise at the stellar boundaries 
     due to discontinuities in the fluid variables, these are not expected to converge to zero.
     } \label{fig:idconv}
\end{center}
\end{figure}

We use initial data generated with the program \magstar, part of
the \lorene\  software package. These solutions are
described in~\cite{Bocquet:1995je} and are rigidly rotating, magnetized
neutron stars. They are generated as fully consistent
solutions of Einstein's equations as opposed to taking nonmagnetized
stars and adding a seed magnetic field without re-solving the constraints.
The initial magnetic field is dipolar and aligned with the rotation
axis, produced by a current function $f(A_{\phi})={\rm constant}$ where $A_{\phi}$ is
the toroidal component of the electromagnetic potential vector.

To get a handle on the solutions generated, we first turn off the
magnetic field and compute the ``usual" two-parameter solution 
space as described in~\cite{cst94}. Using the terminology
of~\cite{Bocquet:1995je}, we compute solutions based on the two parameters
of central enthalpy and frequency. Examination of Eq.~(13) of~\cite{Bocquet:1995je} shows that the log-central enthalpy $H_c$ is related to the entropy used here ($h_e$ as defined in Eq.~(\ref{eq:enthalpy})) by
\begin{equation}
H_c = \ln \left( h_e \right) + C
\end{equation}
where $C$ is constructed by physical constants and $h_e$ is evaluated at the center of the star.

These nonmagnetized solutions are
diagrammed in a plot of total gravitational mass versus central enthalpy
in Fig.~\ref{fig:solutionspace}. The lower curve shows the static limit
for stars which are not rotating. The upper curve is the
mass shedding limit, represented by the largest frequency for which {\tt Magstar}
returned a solution. These curves serve as the upper and lower bounds
on the solution space of stationary, unmagnetized stars.
It should be noted that $\magstar$ generates only rigidly rotating stars,
and therefore the evolutions are far from the fast rotating regime expected
to excite large and growing nonaxisymmetric modes.

We also compute and show a sequence of stars at constant baryon mass.
Such sequences are important because real stars are expected to conserve
baryon mass as they evolve and thus the sequences are expected to
approximate their evolution. Furthermore, along such sequences it has
been shown that there is a stability change at the minimum in the
angular momentum~\cite{cst94}. Looking at the inset of
Fig.~\ref{fig:solutionspace}, one therefore expects the solutions on the
right to be unstable while those on the left should remain stable.

Note that the addition of a non-vanishing magnetic field adds another dimension
to this diagram although the effect of the magnetic field on the initial data
is not so dramatic.
Certainly there are many ways in which to ``add'' a magnetic
field to a stellar solution. In~\cite{Bocquet:1995je} a non-vanishing
current is assumed and a solution is obtained with the same baryon mass but now with nonvanishing 
magnetic field.
In particular, solutions at the mass shedding limit with no magnetic field are shown
in Fig.~\ref{fig:solutionspace} by open triangles. Almost indistinguishable from these
solutions are the those magnetized such that the radial magnetic field at
the pole of the star is $1000~{\rm gigatesla}$~(\gigatesla) or $10^{16} G$.
shown with open circles. 



We consider different perturbations. To perturb the pressure, we
decrease it according to real parameters $A_p$ and $m$ in terms of
the unperturbed pressure $p_0$ depending on the azimuthal angle $\varphi$
\begin{equation}
p = p_0 \left[ 1 - A_p \sin^2 \left( m\varphi \right)  \right].
\label{eq:perturbp}
\end{equation}
We also consider perturbations to the rest mass density decreasing it everywhere
by a fraction $A_\rho$
\begin{equation}
\rho = \rho_0 \left[ 1 - A_\rho \right].
\label{eq:perturbrho}
\end{equation}

We verify the initial data in our evolution
code in  a number of ways. We examine convergence of the data to a unique
solution which solves the constraints. 
In Fig.~\ref{fig:idconv} we show the residual
of the Hamiltonian and $y$-component of the momentum constraint for
three different resolutions. The Hamiltonian constraint is a good
measure of the fidelity of the solution to the Einstein equations, and,
as is clear from the figure, its residual decreases rapidly with
resolution. Furthermore, we confirm that the divergence of the magnetic
field is around machine precision.

\section{Results}
\label{sec:results}

\begin{figure}
\begin{center}
\epsfig{file=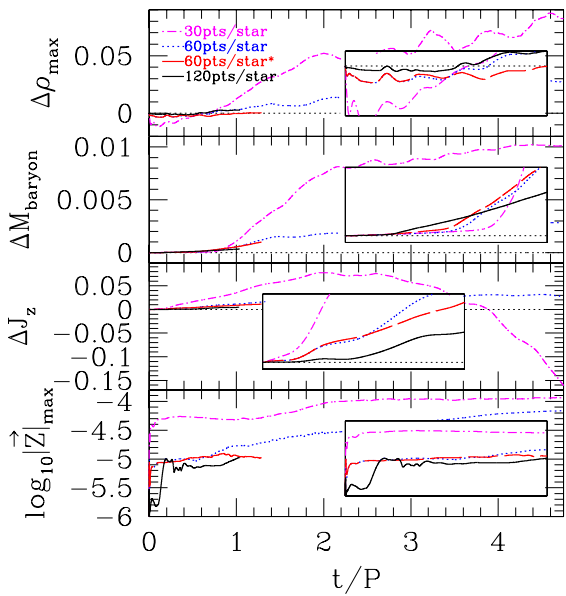,height=9.0cm}
\caption{Convergence results for the evolution of a stable, nonmagnetized, rotating star
     (the particular initial solution is shown in Fig.~\ref{fig:solutionspace} as a
     green solid square). Three FMR evolutions are shown, each with a multiple of the
     coarsest resolution resolving the equator of the star with:
     (magenta, dot-dashed) 30 points/star,
     (blue, dotted) 60 points/star,
     and (black, solid) 120 points/star. 
     Also shown is a run with the same stellar resolution of
     (red, long-dashed) 60 points/star but with
     a coarse grid that extends twice as far in all directions. 
     The two highest resolution runs were terminated early because of the computational 
     cost, not because of any robustness problems.
     The insets show the same data but in finer detail for the first rotational period.
     The {\bf top} frame shows the fractional change in maximum density versus rotational
     period. With increasing resolution the solution better approximates a stationary
     solution.
     The {\bf upper middle} frame shows the fractional change in the baryon mass,
     Although our scheme, using vertex centered AMR, an atmosphere, and full general
     relativity with sources, is not strictly conservative, the plot shows that
     deviations from strict conservation are small and decrease with more resolution.
     The {\bf lower middle} frame shows the fractional change in integrated 
     angular momentum. Again, the code demonstrates convergence to conservation.
     The {\bf bottom} frame shows the maximum of the norm of the $Z^a$ constraint residuals
     as a function of time which also converge.
     } \label{fig:stable}
\end{center}
\end{figure}

\noindent {\bf Stable, Rotating Star Test:}\\
Before addressing the effects of the magnetic field, we verify that
the code reproduces the expected behavior as described, for example,
in~\cite{Font2002}. First, we consider evolutions of stable, rotating
stars and find that the code evolves such a star as long as desired
while maintaining a stationary solution. This a demanding
test as it depends on the balance between gravitational
and hydrostatic forces in a rotating configuration 
with both a non-conforming grid
and variables not adapted to the symmetries of the problem.

One example of the the behavior of the numerical solution is shown in Fig.~\ref{fig:stable}. 
As apparent in the top frame of the figure, the fractional change in
the maximum of the density oscillates with a slow overall increase.
This oscillation is characteristic of quasinormal ringing of the star
excited by inherent numerical error. However, the figure shows data
for a number of resolutions, and the trend as resolution improves
is toward a flatter curve. This trend is particularly apparent
in the inset which shows just the first rotational period.
This behavior suggests that the code is converging
to the continuum solution.

In addition to the runs with varying resolution, Fig.~\ref{fig:stable}
shows another evolution with a coarse level extending twice as far but
with fine levels identical to the medium resolution run just discussed.
Generally, the results are the same as for the non-extended
domain, indicating little effect from the boundary for the the
first half-period.

Although the fluid scheme used here is not strictly conservative
because of (i) AMR boundaries for our vertex centered scheme, (ii) our
outer boundary treatment, and (iii) our use of a fluid floor, 
Fig.~\ref{fig:stable}
shows that the fractional change in the volume-integrated
baryon mass remains constant to a high degree. Similarly,
the integrated angular momentum of the fluid converges to conservation.

Finally, the bottom
frame of Fig.~\ref{fig:stable}
shows the maximum value of the constraint violation.
These values increase with time as the numerical error accumulates, but
higher resolution runs demonstrate less violation though it saturates
due to the intrinsic error of the initial and boundary data.

\begin{figure}
\begin{center}
\epsfig{file=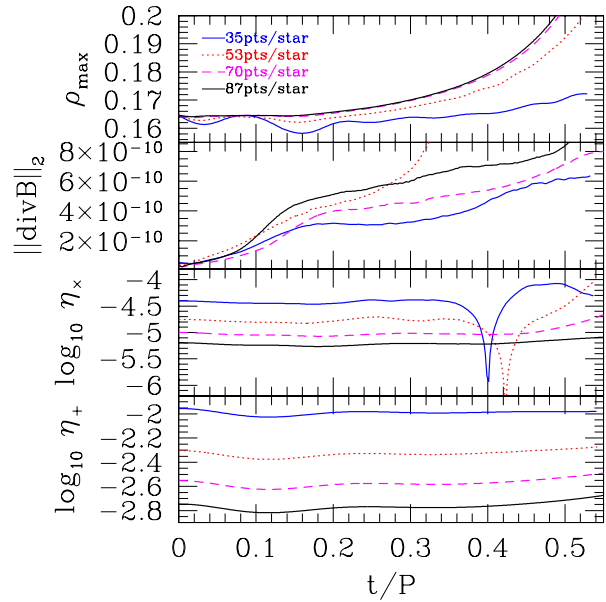,height=9.0cm}
\caption{Collapse to black hole of an unstable, unperturbed, magnetized star for multiple of a base
         resolution.
         The initial data is an unperturbed \magstar\ solution with central enthalpy $H_c = 0.8$
         rotating at a frequency $f=835 Hz$ and with polar magnetic field of $1000~\gigatesla=10^{16}~G$.
         The {\bf top} frame shows the maximum density which increases with time as the star collapses.
         The {\bf upper middle} frame shows a norm of the divergence of the magnetic field.
         The {\bf lower} frames show the distortion parameters of the $D$ field as functions of time.
     } \label{fig:longterm}
\end{center}
\end{figure}

~\\
\noindent {\bf Unstable, Rotating, Magnetized Star Test:}\\
Similarly, we evolve a rotating, magnetized  star located on the unstable side.
These stars, perturbed by inherent numerical error, collapse to black holes.
We see no evidence of significant disk 
formation 
(as in e.g.~\cite{Baumgarte:2002vu,Baiotti:2004wn,Hawke:2005mg,Duez:2005cj,Stephens:2008hu}).  
Notice that even though we do not impose any type of symmetry,
no asymmetric, unstable modes are observed. This behavior is consistent with previous
work which studied the possible onset of axisymmetric 
instabilities~\cite{Manca:2007ca}. These previous studies found that such instabilities require 
high rotation rates characterized by $T/W > 0.25$ in contrast to those studied here for which
$T/W \leq 0.1$.
Consequently, we see no evidence for deformations of the star as would be apparent by monitoring
 the distortion parameters moving away from zero.

In Fig.~\ref{fig:longterm}, we show an example of such an evolution at
three successively finer resolutions. Because the star is collapsing,
the maximum density increases dramatically. The magnetic field remains
essentially poloidal throughout the collapse and  its maximum magnitude 
increases due to the resulting compression of the field lines. 
Despite this increase, the magnetic field plays no 
important role in the collapse since its associated pressure is still
several orders of magnitude smaller than the fluid pressure.
While we do observe an increase in the norm of the divergence,
the growth is not particularly fast and the divergence remains small in
absolute terms and relative to the magnetic field.

\begin{figure}
\begin{center}
\epsfig{file=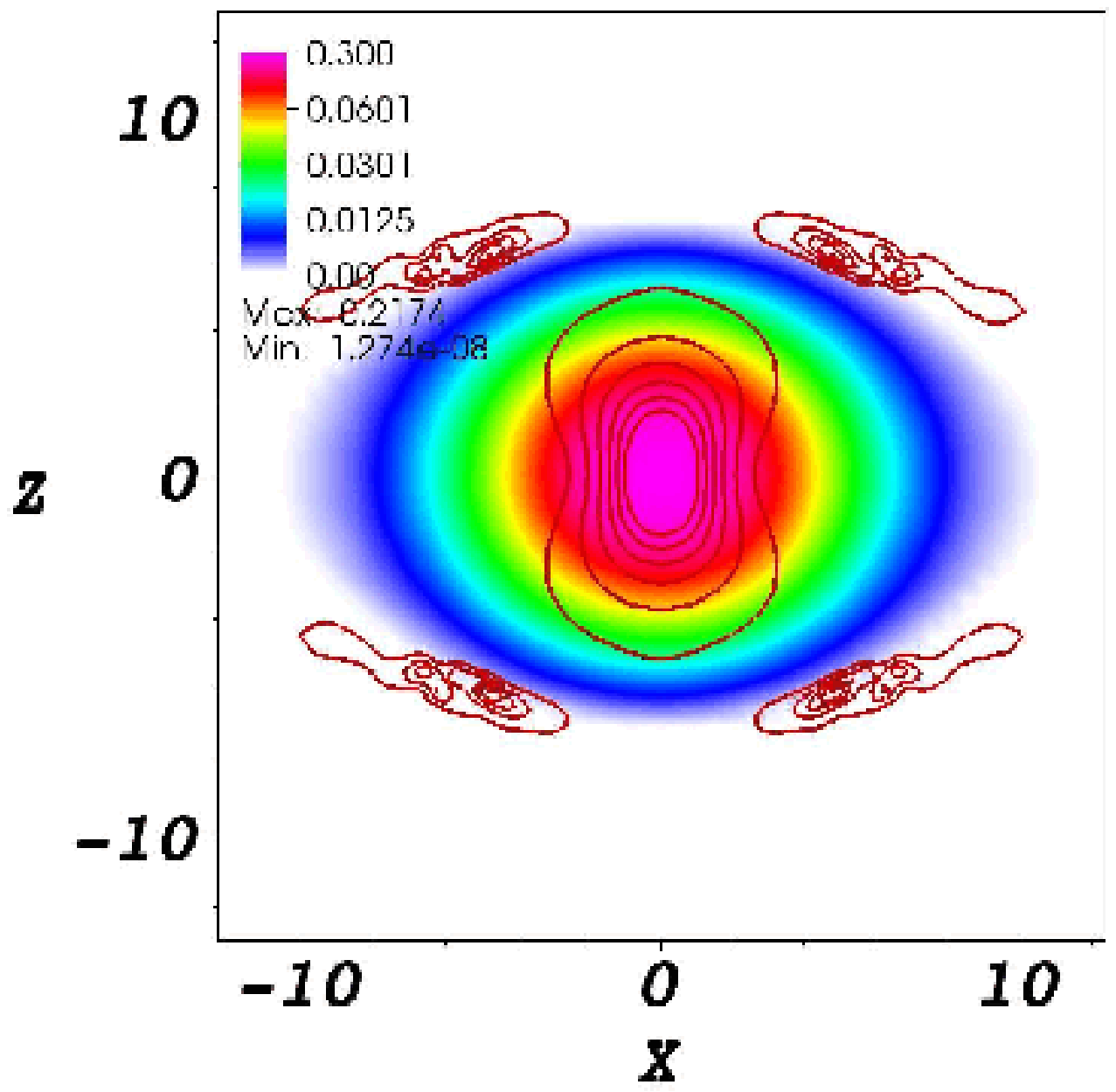,height=4.2cm}
\epsfig{file=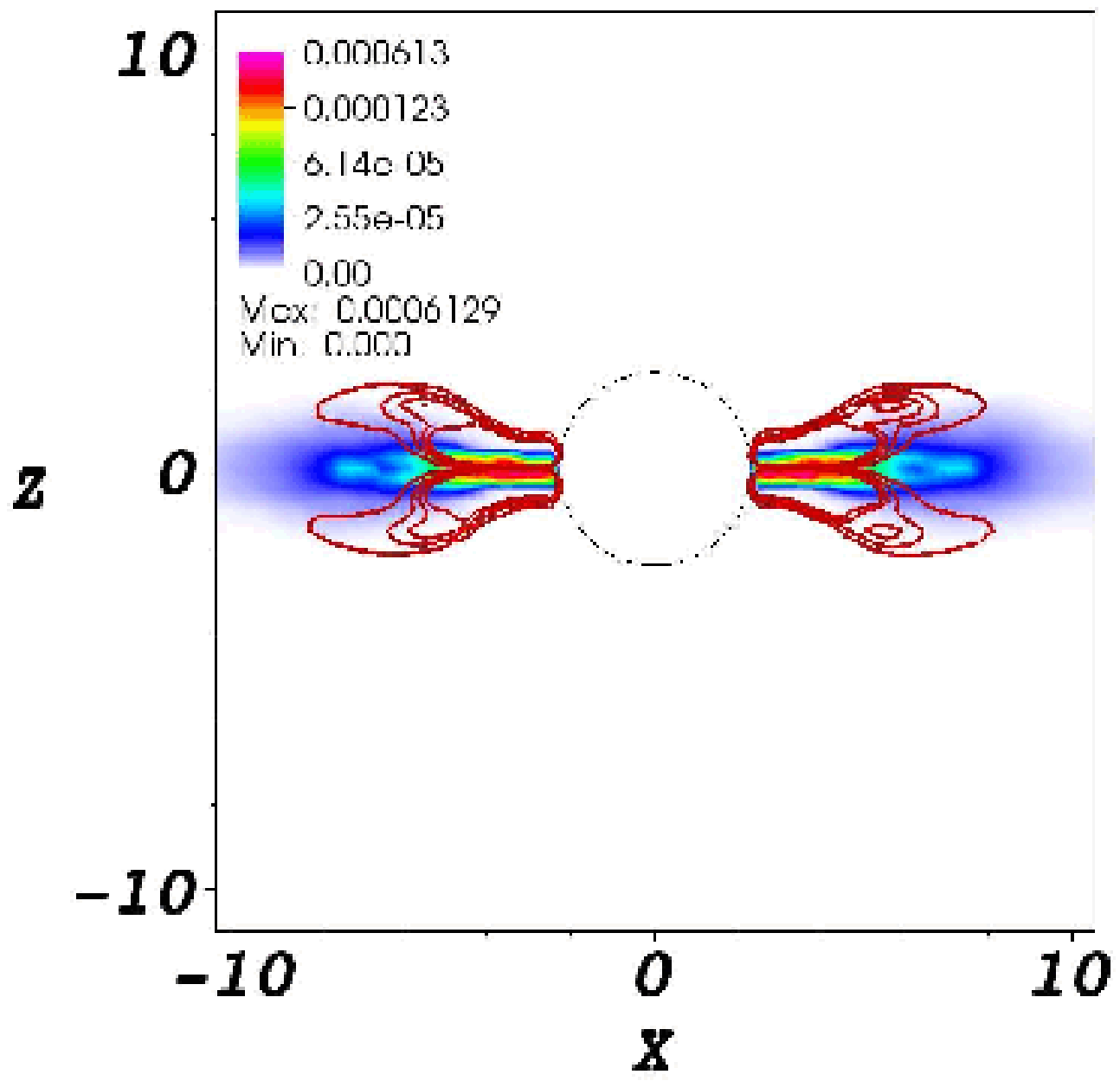,height=4.2cm}
\epsfig{file=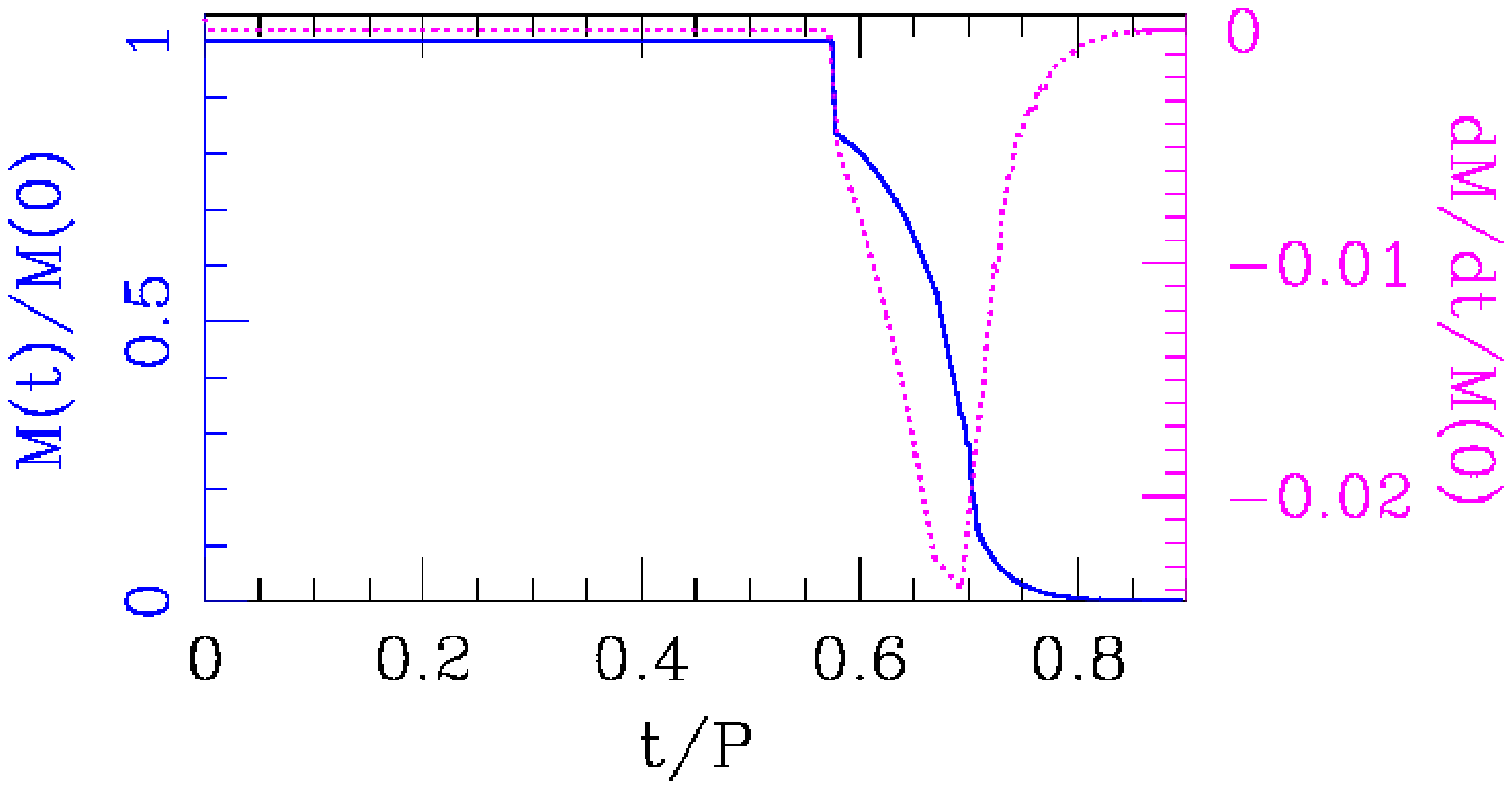,height=4.2cm}
\caption{Density $\rho_0$ and magnetic field strength  
on the $y=0$ plane for the same star as shown in Fig.~\ref{fig:longterm}.
The density is shown with respect to 
the colormap while the contours denote the magnetic field,
which is equally spaced from $0$ to $5\times 10^{15}$G.
%
%
The top left corresponds to $t=0.5P$
while the the top right to $t=0.9P$ (the circle shown
represents the apparent horizon.) 
The bottom plot illustrates the normalized mass as a function of time, along
with its rate of change.
     } \label{fig:excision}
\end{center}
\end{figure}

%
%
%
Fig.~\ref{fig:excision} displays two snapshots of the density and magnetic
field strength of the collapsing star tested in Fig.~\ref{fig:longterm}.
The first one illustrates a stage during the collapse before an 
apparent horizon forms. The second one shows the behavior after  
an apparent horizon is found and its interior excised.
The apparent horizon appears at $t \simeq 0.6\,P$, when the maximum of the density is 
$\rho_{max}=0.271$ and the minimum of the lapse is $\alpha_{min}=0.2$.

\begin{figure}
\begin{center}
\epsfig{file=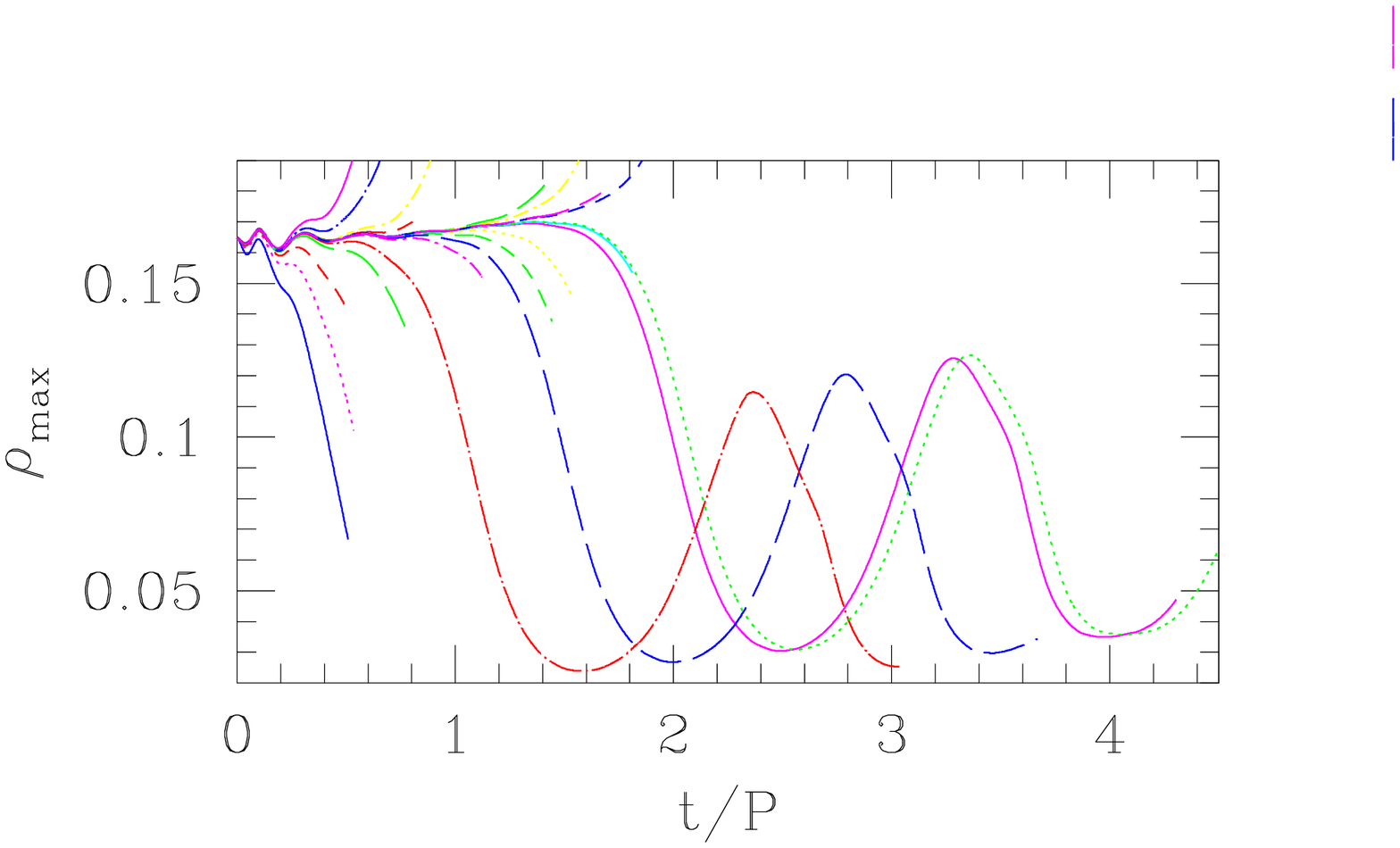,height=9.0cm}
\vspace{-1.5in}
\caption{The maximum density $\rho_{\rm max}$ as a function of time
     for an unstable, perturbed ($m=3$ in Eq.~(\ref{eq:perturbp})), non-magnetic, rotating star.
     The (unperturbed) initial star has central enthalpy $0.8$ and rotates
     at the mass shedding limit. Solutions resulting from 
     tuning the amplitude of the perturbation $A_p$  to roughly one part in $10^6$ show the two disparate outcomes~(as
     discussed in~\cite{Font2002}): 
     collapse to black hole or 
     violent oscillations about a stable
     stellar solution with equivalent mass. With more tuning, the star resembles the initial,
     unstable solution for a longer time. 
     } \label{fig:unstablenob}
\end{center}
\end{figure}
~\\
\noindent {\bf Perturbations of Unstable Stars}:\\
Previous work presented in~\cite{Font2002} argues that, in general, unstable stars should either
collapse to a black hole or expand and oscillate about a stable stellar solution.
Seeking to duplicate this behavior, we perturb an unstable star. Indeed,
as shown in Fig.~\ref{fig:unstablenob}, we find precisely the behavior described.
If our perturbation increases the pressure sufficiently, we find a star which
expands and oscillates about some other, presumably stable solution. However,
if we choose a perturbation which barely increases the pressure, the star
collapses to a black hole.

However, given the interest in black hole critical phenomena (see \cite{Choptuik:1992jv,Gundlach:2007gc}) over the past couple
of decades, we study in detail the separation between these two behaviors.
That is, we continue to adjust $A_p$ in Eq.~(\ref{eq:perturbp})
searching for a value $A^*_p$ above which one finds black hole formation,
and below which one finds an expanding solution (the way we have parameterized
the pressure perturbation, $A_p^*<0$).

As apparent from Fig.~\ref{fig:unstablenob}, the more one continues this
tuning, the longer the unstable star survives. This type of tuning is
reminiscent of a similar analysis of an unstable, irregular static
solution~\cite{Choptuik:1997rt}. What these results suggest is that
the unstable solution (i)~sits at the threshold of black hole formation with
(ii)~a single unstable mode. That small perturbations about the
solution send it either to collapse or expansion suggest that it sits
at the threshold. Furthermore that a single parameter is sufficient to
stabilize the solution suggests that there is a single unstable mode. In contrast, sometimes one
can tune multiple parameters to find a threshold solution~\cite{Liebling:1998xu}.

If  these suggestions hold up, these would suggest that at least some of
these unstable solutions might serve as Type I critical solutions. Indeed,
previous work~\cite{NoblePHD} perturbed stable TOV stars in spherical 
symmetry and found unstable TOV stars at criticality in Type I collapse.
In that work, they were able to achieve phenomenal resolution and tuning.
In contrast, while they perturbed a self-consistent {\em stable} solution
and saw the tuned evolution driven to the {\em unstable} branch of solutions,
here we begin with the unstable solution and perturb around it.
Here, we have only been able to tune to about one part in a million, being
prevented from tuning further because successive evolutions stop the trend towards
longer lived solutions. That such searches are prevented from continuing
might indicate some new phenomena, or, more likely, that boundary effects and numerical
error begin to spoil the threshold behavior.\\

We have looked for a scaling law in the survival times of these tuned unstable
stars. To the extent that this rough tuning is representative of the overall
behavior, we find that the different solutions appear to scale as expected. 
However, because our
searches have terminated so far from criticality, we cannot have much confidence in
a precise scaling relationship.
There has been recent work in the axisymmetric collision of 
neutron stars~\cite{Jin:2006gm,Wan:2008th} which appears to demonstrate 
the same type of critical behavior as observed in~\cite{NoblePHD}.

\begin{figure}
\begin{center}
\epsfig{file=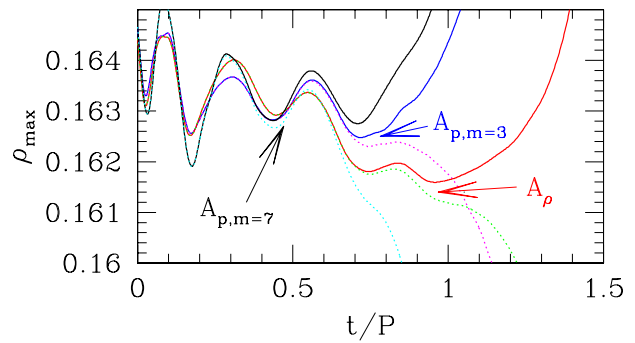,height=9.0cm}
\vspace{-1.5in}
\caption{Results of three critical searches for a magnetized, rotating star with
     central enthalpy of $0.8$ and a radial magnetic field at the pole of $1,000 GT$.
     One search varied the amplitude of the
     pressure perturbation $A_p$ with $m=7$ (black and cyan),
     another with $m=3$ (blue and magenta), and another 
     varied the density perturbation $A_\rho$. In solid line are shown
     the super-critical evolutions which collapse to black holes, while sub-critical
     solutions for the two searches are shown with dotted lines.
     The search
     over pressure with $m=7$ achieved tuning to about one part in $10^3$,
     that with $m=3$, one part in $10^4$,  and that over
     density achieved about one part in $10^5$.
     } \label{fig:unstablewb}
\end{center}
\end{figure}

We find the same type of behavior about a magnetized star
as shown in Fig.~\ref{fig:unstablewb}. Here
we have carried out three searches by varying a different
parameter all perturbing the same solution. The figure makes apparent the same ringing for
all three families, although the amplitude varies across the different tuning families.
It seems reasonable to take the results of these tunings as further evidence that only 
a single mode is unstable since if there were more unstable modes, these different tunings
would produce solutions more varied from each other.
We have begun to look at the geometry of the purported unstable mode
by looking at the difference $\Delta = \rho_0(t)-\rho_0(0)$ at
late times for near-critical evolutions. These
calculations indicate the the mode is likely axisymmetric with
differences between $\Delta$ on the $x=0$ plane and on the $y=0$ plane
being at about the $10\%$ level.

\section{Conclusions}
\label{sec:conclusions}

We study the evolution of rotating stellar configurations and
examine different phenomenology related to stability and magnetic
field influence in their dynamical behavior. The stars are constructed 
assuming a polytropic equation of state using the code {\tt Magstar}
which is part of the publicly available {\sc Lorene} package.
%
We present several studies which indicate our code reliably
and robustly evolves astrophysically relevant scenarios including
magnetic field effects.  
%
%
We study the dynamical effect of perturbations of stars on the unstable branch of solutions.
We find evidence that these unstable solutions may play a similar
role as the unstable TOV stars play in spherically symmetric
evolutions as studied in~\cite{NoblePHD}. This is significant because it suggests that the
addition of angular momentum, magnetic field, and three dimensions do not 
allow for a multitude of unstable modes. Needless to say, the
phenomena associated with the threshold of gravitational collapse
merit further study. 

Beyond the studies considered in this work, further interesting phenomenology
to consider include the impact of magnetic fields in the stability of the star,
a thorough analysis of the possible critical phenomena observed and differences
due to more realistic equations of state. The investigation of such scenarios
will be presented elsewhere.

%
%
\begin{acknowledgments}
We would like to thank J. Novak for his gracious assistance with \magstar\ and \lorene.
SLL thanks Matthew Choptuik and Scott Noble for valuable discussions concerning the
apparent critical behavior. We also would like to thank Matthew Anderson, Eric Hirschmann,
Miguel Megevand, Patrick Motl, Joel Tohline and Travis Garret for comments and suggestions.
SLL, DN and CP thank the Perimeter Institute for Theoretical Physics for hospitality where
parts of this work were completed.
This work was supported by the National Science Foundation under grants
PHY-0803629 and PHY-0653369 to Louisiana State 
University, CCF-0832966 and PHY-0803615
to Brigham Young University, and CCF-0833090, PHY-0803624 to Long Island University.
and NSERC through a Discovery Grant.  Research at Perimeter Institute is
supported through Industry Canada and by the Province of Ontario
through the Ministry of Research \& Innovation.
This research was also supported in part by the National Science Foundation
through TeraGrid resources provided by SDSC under allocation award PHY-040027.
In addition to TeraGrid resources, we have employed clusters belonging
to the Louisiana Optical Network Initiative (LONI), and clusters
at LSU and BYU.
\end{acknowledgments}

%
%
\bibliography{./magstar}
\bibliographystyle{apsrev}

%
%
\end{document}